\documentclass[prd,onecolumn,floats,floatfix,showpacs,nofootinbib,11pt,notitlepage]{revtex4-1}
\usepackage{graphicx}
\usepackage{dcolumn}
\usepackage{bm}
\usepackage{graphics}
\usepackage{slashed}
\usepackage{amssymb}
\usepackage{natbib}
\usepackage{amsmath}
\usepackage{url}
\usepackage[usenames,dvipsnames,svgnames,table]{xcolor}
\newcommand{\bea}{\begin{eqnarray}}
\newcommand{\ena}{\end{eqnarray}}
\newcommand{\bean}{\begin{eqnarray*}}
\newcommand{\enan}{\end{eqnarray*}}

\begin{document}

\title{Non-linear electrodynamics non-minimally coupled to gravity:\\ symmetric-hyperbolicity and causal structure.}

\author{\'Erico Goulart}
\address{Federal University of S\~ao Jo\~ao d'El Rei, C.A.P. Rod.: MG 443, KM 7, CEP-36420-000, Ouro Branco, MG, Brazil}

\author{Santiago Esteban Perez Bergliaffa}
\address{Departamento de F\'isica Te\'orica, Instituto de F\'isica, Universidade do Estado de Rio de Janeiro, 20550-013, Rio de Janeiro, Brazil}

\begin{abstract}

It is shown here that symmetric hyperbolicity, which guarantees well-posedness,  
leads to a set of two inequalities for matrices whose elements are determined by a given theory. As a part of the calculation, carried out in a mostly-covariant formalism, the general form for the symmetrizer, valid for 
a general Lagrangian theory, was obtained.
When applied to nonlinear electromagnetism linearly coupled to curvature, the inequalities 
lead to strong constraints on the relevant quantities, which were illustrated with applications to particular cases. The examples show that 
non-linearity leads to constraints on the field intensities, and non-minimal coupling
imposes restrictions on quantities associated to curvature. 

\end{abstract}
 
\maketitle

%%%%%%%%%%%%%%%%%%%%%%%%%%%%%%%%%%%%%%%%%%%%%%%%%%%%%%%%%%%%%%
%%%%%%%%%%%%%%%%%%%%%%%%%%%%%%%%%%%%%%%%%%%%%%%%%%%%%%%%%%%%%%
%%%%%%%%%%%%%%%%%%%%%%%%%%%%%%%%%%%%%%%%%%%%%%%%%%%%%%%%%%%%%%
%%%%%%%%%%%%%%%%%%%%%%%%%%%%%%%%%%%%%%%%%%%%%%%%%%%%%%%%%%%%%%
  \section{Introduction}

Well-posedness of the initial value problem stands out as a basic requirement to be satisfied by any relativistic field theory. Broadly speaking, it implies that solutions for a given problem exist, are unique, and depend continuously on the initial data. 
%\footnote{\santiago{Shock-free propagation (see for instance \cite{McCarthy1999}) is another desirable property, which is implied by well-posedness (??)}}.
% An immediate question with the utmost relevance
%is whether the nonlinear partial differential equations (PDEs) admit a well-posed Cauchy problem. 
Hence, well-posedness is at the roots of physics, for
it amounts to the predictability power of a given theory.
While it is difficult to 
decide whether a given non-linear theory  has a well-posed initial value problem, 
a necessary and sufficient condition for well-posedness around a given solution is 
that 
all the linearised
problems obtained by linearising near such a solution are well-posed, see for instance \cite{kreiss,Papallo2017}.

In the linearized regime, well-possedness
means that the equation of motion is hyperbolic.
At least three notions of hyperbolicity can be distinguished
\cite{kreiss,Beig2004}:
(a) weak hyperbolicity, in which all the roots of the characteristic equation are
real, (b) strong hyperbolicity implies   
that there is an energy estimate that sets a bound for the energy of a solution at a given time in terms of the initial energy \footnote{In fact, strong hyperbolicity  
is a necessary and sufficient condition for the
initial-value problem to be well-posed.}, and (c) symmetric hyperbolicity, which is a sufficient  condition for well-posedness 
\cite{Sarbach2012}.
It follows that symmetric hyperbolicity implies 
strong hyperbolicity, which in turn implies weak hyperbolicity.

There are many examples in which the requirement of some kind of hyperbolicity imposes severe restrictions on the Lagrangian of a given theory. For instance, as shown in 
    \cite{Papallo2017},
    the equations of motion
    of Lovelock'ss theory are always weakly hyperbolic for weak fields but not strongly hyperbolic
    in a generic weak-field background. The well-posedness of Horneski theory has been analyzed in 
    \cite{Papallo2017,Papallo2017b}. As shown in the latter reference, 
 the most general Horndeski theory which is strongly hyperbolic for weak fields in a generalized harmonic gauge is simply  \emph{k-essence} coupled to Einstein's gravity (see also \cite{Bernard2019}).
The well-posedness of 
scalar-tensor effective field theory
was studied in \cite{Kovacs2020},
where it was shown that 
the
equations of motion are strongly hyperbolic at weak coupling.

We would like to explore here the constraints imposed by the requirement of well-posedness in the case of nonlinear electromagnetic theories coupled to gravity. Nonlinear electromagnetism has been widely studied in several contexts. A non-exhaustive list of applications and references includes black holes
\cite{Moreno2002,Breton2015,Toshmatov2018,Magos2020, Falciano2021}, astrophysics 
\cite{Heyl2003,Harding2006,Turolla2015}
and cosmology 
\cite{Novello2003,Novello2006,Campanelli2007,Vollick2008,Kruglov2015}.
There are also several articles devoted to different aspects of the  propagation of perturbations in nonlinear electromagnetic theories, such as 
\cite{Gutierrez1981, Deser1998,Goulart2009, Perlick2018}. The matter of (symmetric) hyperbolicity
for nonlinear electromagnetism minimally coupled to gravity
was analyzed in a flat 
spacetime background in \cite{Abalos2015}, while
the hyperbolicity of Maxwell’s equations with a local (and possibly nonlinear) constitutive law in flat spacetime 
was considered in
\cite{Perlick2010}.
Our aim here is to determine the restrictions that follow from the requirement of symmetric hyperbolicity 
on nonlinear electromagnetism non-minimally coupled to gravity, several aspects of which have been studied in detail 
in \cite{Balakin2005, Balakin2007,Balakin2010,Dereli2011}. 
We shall restrict to couplings linear in the curvature, since higher-order couplings produce higher-than-second order equations for the gravitational field  \cite{Balakin2005}.

As mentioned above, a sufficient condition for well-posedness to hold is that the system under study admits a symmetric-hyperbolic representation. The theory of \textit{first-order symmetric hyperbolic} systems, originally due to Friedrichs \cite{Friedrichs1954}, has been extensively developed (see for instance \cite{Beig2004} and references therein).
To study the evolution of a system we shall adopt here the modern geometric approach to the subject outlined by Geroch in \cite{Hall1996},
in which covariance is kept during most of the calculation, 
instead of using a 3+1 decomposition of spacetime (as for instance in \cite{Perlick2010}) \footnote{For a list of other approaches to the analysis of the evolution of minimally coupled electromagnetism, see \cite{Abalos2015}.}.
We shall see
that such approach leads to a general form for the symmetrizer, and to conditions for symmetric hyperbolicity that are easier to evaluate than those for other types of hyperbolicity.

 The structure of the paper is as follows. In Section \ref{eoms}
 the basic notation 
 used in the equations of motion is presented.
 Symmetric hyperbolicity, the related concept of symmetrizer, and the role of the constraints
are analyzed in Section \ref{symmhyp}. 
  The equations defining the characteristic cones will be deduced in 
  Section \ref{cones}.
  Our main result, namely the explicit form of the matrices that are needed to investigate the symmetric hyperbolicity of any nonlinear electromagnetic theory non-minimally coupled to gravity can be found in Section \ref{result}. In Section \ref{applic}, some examples of the restrictions imposed by symmetric hyperbolicity are presented for different theories. Our closing remarks are presented in Section \ref{concl}.
  
 \section{Lagrangians and equations of motion}
\label{eoms}
To begin with, let $\mathcal{M}$ denote a smooth four-dimensional spacetime with a Lorentzian metric $\boldsymbol{g}$ of signature $(+,-,-,-)$. For the sake of concreteness we assume $\mathcal{M}$ to be also oriented and globally hyperbolic i.e. $M\cong\mathbb{R}\times\Sigma$, with $\Sigma$ a codimension-$1$ hypersurface. Sticking to the conventions $R_{ab}\equiv R^{c}_{\phantom a acb}$ and $[ab]\equiv ab-ba$,  the canonical decomposition of the Riemman tensor into its irreducible 
%factors 
parts reads
%\begin{equation}
%R^{ab}_{\phantom a\phantom a cd}=W^{ab}_{\phantom a\phantom a cd}+E^{ab}_{\phantom a\phantom a cd}+S^{ab}_{\phantom a\phantom a cd}    
%\end{equation}
\begin{equation}
R^{ab}_{\phantom a\phantom a cd}=W^{ab}_{\phantom a\phantom a cd}+\frac{1}{2}\delta^{[a}_{\phantom a [c}S^{b]}_{\phantom a d]}+\frac{R}{12}g^{ab}_{\phantom a\phantom a cd},
\end{equation}
where $W_{abcd}$ is the Weyl conformal tensor, $S_{ab}$ is the traceless part of the Ricci tensor, $R$ is the scalar curvature and $g_{abcd}=g_{a[c}g_{d]b}$ is the Kulkarni-Nomizu product of the metric with itself. Clearly, each factor in the decomposition has the same algebraic symmetries as the full Riemann tensor, that is
\begin{equation}\label{sym1}
R_{abcd}=-R_{bacd}=-R_{abdc},\quad\quad\quad R_{abcd}=R_{cdab},
\end{equation}
\begin{equation}\label{sym2}  R_{abcd}+R_{acdb}+R_{adbc}=0.
\end{equation}  
An arbitrary rank four covariant tensor satisfying Eqs. (\ref{sym1}) has 21 independent components and is often referred to as a double symmetric $(2,2)$-form (the skew pairs can be interchanged). If, in addition, the tensor satisfies Eq. (\ref{sym2}), it is called an algebraic curvature tensor (see for instance \cite{Besse1987}). %Such tensors naturally live in the bundle $S^{2}(\Lambda^{2}\mathcal{M})$, the second symmetric power of the second exterior power of the cotangent bundle.

If we are given a double symmetric $(2,2)$-form on $\mathcal{M}$, say $\chi$, we may 
%produce 
obtain its left-dual $\star\chi$ and right-dual $\chi\star$ by
\begin{equation}
\star \chi_{abcd}=\frac{1}{2}\varepsilon_{ab}^{\phantom a\phantom a pq}\chi_{pqcd},\quad\quad\quad \chi_{abcd}\star=\frac{1}{2}\varepsilon^{pq}_{\phantom a\phantom a cd}\chi_{abpq},
\end{equation}
where $\varepsilon_{abcd}$ is the Levi-Civita tensor with $\varepsilon_{0123}=\sqrt{-g}$. Notice that the place of the $\star$ indicates the pair of skew indices which are Hodge-dualized. A direct consequence is that
\begin{equation}
\star\star\chi_{abcd}=-\chi_{abcd},\quad\quad\quad\chi_{abcd}\star\star=-\chi_{abcd}. 
\end{equation}
We aim at constructing 
Lagrangians
describing the coupling of gravity with the electromagnetic field 
using invariants that are at most linear in the curvature
\footnote{As discussed in \cite{Balakin2005}, higher-order couplings lead to 
equations of motion with derivatives of the metric higher than two.}, and respecting the $U(1)$ gauge invariance of electromagnetism.
Recalling that every algebraic curvature tensor satisfy the Ruse-Lanczos identity,
we may construct  the following independent rank four tensors (see the Appendix for details)\\
\begin{eqnarray}\label{1}
^{(1)}\chi^{ab}_{\phantom a\phantom a cd}\equiv g^{ab}_{\phantom a\phantom a cd},\quad\quad\quad ^{(2)}\chi^{ab}_{\phantom a\phantom a cd}\equiv \varepsilon^{ab}_{\phantom a\phantom a cd},
\end{eqnarray}
\begin{eqnarray}\label{2}
^{(3)}\chi^{ab}_{\phantom a\phantom a cd}\equiv R g^{ab}_{\phantom a\phantom a cd},\quad\quad\quad ^{(4)}\chi^{ab}_{\phantom a\phantom a cd}\equiv R\varepsilon^{ab}_{\phantom a\phantom a cd},
\end{eqnarray}
\begin{eqnarray}\label{3}
^{(5)}\chi^{ab}_{\phantom a\phantom a cd}\equiv W^{ab}_{\phantom a\phantom a cd} ,\quad\quad\quad ^{(6)}\chi^{ab}_{\phantom a\phantom a cd}\equiv \star W^{ab}_{\phantom a\phantom a cd},
\end{eqnarray}
\begin{eqnarray}\label{4}
^{(7)}\chi^{ab}_{\phantom a\phantom a cd}\equiv \delta^{[a}_{\phantom a [c}S^{b]}_{\phantom a d]}. 
\end{eqnarray}
Notice that ${\phantom a}^{(\Gamma)}\chi_{abcd}$ $(\Gamma=1,2,...,7)$ are double symmetric $(2,2)$-forms by construction and it can be checked that they exhaust the interesting possibilities: the first pair, Eqs. (\ref{1}), involves only algebraic terms in the metric, while the remaining Eqs. (\ref{2})-(\ref{4}) contain at most linear terms in the curvature.

Let $F_{ab}=\partial_{[a}A_{b]}$ denote a 
%probe 
test
electromagnetic field propagating on $\mathcal{M}$. We shall
focus on the propagation properties of such a field on a fixed gravitational background. The EM field will be described by 
%consider a
the gauge-invariant action functional of the form
\begin{equation}\label{action}
S=\frac{1}{4}\int d^{4}x\sqrt{-g}\ \mathcal{L}(I^{1},I^{2},...,I^{7}),
\end{equation}
where the factor $1/4$ is introduced for future convenience and the Lagrangian density is taken as an arbitrary smooth function of the following scalars
\begin{equation}\label{inddef}
I^{\Gamma}\equiv {\phantom a }^{(\Gamma)}H^{ab}F_{ab},\quad{\rm where}\quad\quad {\phantom a }^{(\Gamma)}H^{ab}\equiv \frac{1}{2}{\phantom a }^{(\Gamma)}\chi^{ab}_{\phantom a\phantom a cd}F^{cd}.
\end{equation}
In analogy with electrodynamics in material media, we call ${\phantom a }^{(\Gamma)}H^{ab}$ the $\Gamma$-th \textit{induction tensor} and ${\phantom a }^{(\Gamma)}\chi^{ab}_{\phantom a\phantom a cd}$ the $\Gamma$-th \textit{constitutive tensor}. Particular instances described by  Eq. (\ref{action}) include Maxwell's theory, minimally-coupled nonlinear electrodynamics, and the three-parameter non-minimal Einstein–Maxwell model originating from QED vacuum polarization in a background gravitational field (see for instance \cite{Drummond1979}), among others. With the conventions presented above, the variation of the action with respect to the 4-potential yields a coupled system of first-order quasi-linear PDEs for the fields, given by
\begin{equation}\label{EOM}
\nabla_{b}H^{ab}=0,\quad\quad\quad \nabla_{b}\star F^{ab}=0.
\end{equation}\\
Here, $\nabla_{a}$ is the covariant derivative compatible with the metric and the \textit{full induction tensor} is defined by
\begin{equation}
H^{ab}\equiv\displaystyle\sum \mathcal{L}_{\Gamma}{\phantom a}^{(\Gamma)}H^{ab},
\end{equation}
with $\mathcal{L}_{\Gamma}\equiv\partial\mathcal{L}/\partial I^{\Gamma}$, for conciseness. Notice that the system defined by Eqs. (\ref{EOM}) is composed of eight equations for only six unknowns. In relevant physical situations, the equations \textit{must} include six dynamical equations 
%(depending on 
(defined with respect to some time coordinate, to be identified later) and two constraints, which are to be imposed on initial data.

\section{Symmetric-hyperbolicity}
\label{symmhyp}
%We shall consider the dynamics of a probe electromagnetic field, $\boldsymbol{F}\in\Omega^{2}(\mathcal{M})$, provided by the equations of motion
%\begin{equation}\label{ql}
%\nabla_{m}\left(X^{ma}_{\phantom a\phantom a\phantom a bc}F^{bc}\right) =0,\quad\quad\quad\nabla_{m}\left(\varepsilon^{ma}_{\phantom a\phantom a\phantom a bc}F^{bc}\right)=0.
%\end{equation}
%Here, $\nabla$ is the covariant derivative compatible with the metric, $\varepsilon_{abcd}$ is the Levi-Civita tensor with $\varepsilon_{0123}=\sqrt{-g}$ and $X_{abcd}$ is a smooth double (2,2)-form, possibly depending on position and on the electromagnetic scalars
%\begin{equation}
%\psi=\frac{1}{2}F_{ab}F^{ab},\quad\quad\quad\phi=\frac{1}{2}\star F_{ab}F^{ab}, \quad\quad\quad \star F_{ab}=\frac{1}{2}\varepsilon_{abcd}F^{cd}.
%\end{equation}
%In general, a double (2,2)-form has 36 independent components. However, for reasons which will be clarified latter on, we assume that it is also symmetrical i.e.,
%\begin{equation}\label{algprop}
%X_{abcd}=X_{cdab}.
%\end{equation}
%This symmetry reduces the number of independent components to 21, which is sufficient for our purposes here. Clearly, Eqs. (\ref{ql}) define a system of first order quasi-linear equations for the fields and, in physical situations, they will include dynamical equations depending on some time coordinate and also the constraints, which must be imposed on initial data.
%propagating in a curved spacetime $M$ described by the action\\
%\begin{equation}\label{action}
%S_{vec}= -\frac{1}{4g^{2}}\int d^{4}x \sqrt{-g}\ X^{abcd}F_{ab}F_{cd},
%\end{equation}\\

%and, 
In order to study whether the equations of motion admit a symmetric-hyperbolic representation, it is convenient to recast the system of equations Eqs. (\ref{EOM}) in a unified manner as
\begin{equation}\label{KF}
K_{A \phantom a \beta}^{\phantom a m}(x,\Phi)\partial_{m}\Phi^{\beta}+J_{A}(x,\Phi)=0,
\end{equation}
where $x\in{\mathcal{M}}$, $K_{A \phantom a \beta}^{\phantom a m} $ is the \textit{principal part} of the PDE and $J_{A}(x,\Phi)$ stands for semi-linear contributions
(whose explicit form is unnecessary for our discussion). Here capital Latin indices $(A=1,...,8)$  stand for the space of multi-tensorial equations, lowercase Latin indices $(m=0,1,2,3)$ stand for space-time indices, and greek indices $(\beta=1,..,6)$ for tensorial unknowns. To start with, we introduce an ordering of the antisymmetric indices to obtain the six possible collective quantities 
\begin{equation}
1\rightarrow (01)\quad 2\rightarrow (02)\quad3\rightarrow (03)\quad 4\rightarrow (32)\quad 5\rightarrow (13)\quad 6\rightarrow (21).
\end{equation}
Making the identification $\Phi^{\beta}\rightarrow F^{bc}$ and performing simple manipulations in Eqs. (\ref{EOM}), the principal part is then written as
\begin{equation}\label{K}
K_{A \phantom a \alpha}^{\phantom a m}=\frac{1}{2}\left(X_{a\phantom a\phantom a bc}^{\phantom a m},\ \varepsilon_{a\phantom a\phantom a bc}^{\phantom a m} \right),
\end{equation}
with
\begin{equation}\label{XL}
X_{abcd}\equiv \sum_{\Gamma} \mathcal{L}_{\Gamma}{\phantom a}^{(\Gamma)}\chi_{abcd}+4\sum_{\Gamma}\sum_{\Lambda}\mathcal{L}_{\Gamma\Lambda}{\phantom a}^{(\Gamma)}H_{ab}{\phantom a}^{(\Lambda)}H_{cd}.
\end{equation}
$X_{abcd}$ consists of
%splits into 
a main term involving only first partial derivatives of the Lagrangian density and a nonlinear term including the Hessian matrix of the latter. Clearly, if the Lagrangian is a linear combination of the invariants $I^\Gamma$, the last term vanishes and linear equations of motion follow. More importantly, $X_{abcd}$ is always a symmetric double (2,2)-form independently on the specific form of the Lagrangian. This is a direct consequence of the symmetries of the constitutive tensors together with the symmetry of the Hessian matrix, {\it i.e.} $\mathcal{L}_{\Gamma\Lambda}=\mathcal{L}_{\Lambda\Gamma}$.

In what follows we shall use the covariant approach for first-order symmetric-hyperbolic systems outlined in \cite{Hall1996}:
%.  Following the latter, 
a \textit{symmetric hyperbolization} of Eqs. (\ref{KF}) means 
that there exist
a smooth symmetrizer $h^{A}_{\phantom a\alpha}$ and a covector field $n_{m}$, such that:\\

\begin{enumerate}
\item{$\hat{K}_{\alpha\phantom a\beta}^{\phantom a m}\equiv h^{A}_{\phantom a\alpha}K_{A \phantom a \beta}^{\phantom a m}$ is symmetric in the indices $\alpha,\ \beta$}.\\
\item{The matrix $\hat{K}_{\alpha\beta}(n)\equiv\hat{K}_{\alpha\phantom a\beta}^{\phantom a m}n_{m}$ is positive-definite.}\\
\end{enumerate}
Roughly speaking, the first statement means that it should be possible to construct from Eqs. (\ref{KF}) a new subsystem of first-order quasi-linear PDEs given by
\begin{equation}
\hat{K}_{\alpha\phantom a\beta}^{\phantom a m}\partial_{m}\Phi^{\beta}+\hat{J}_{\alpha}=0
\end{equation}
with $\hat{J}_{\alpha}\equiv h^{A}_{\phantom a\alpha}J_{A}$. Clearly, the new system contains only evolution equations and its dimension is equivalent to the number of unknown fields. The second statement means that the new system can be solved uniquely for any given set of initial conditions on a hypersurface $\Sigma$ with normal covector $n_{m}$. What are the remaining equations the symmetrizer does not capture? For the system to be consistent they should not be of the evolution type. In other words, they must be satisfied automatically once they are satisfied initially i.e., they should be what we normally call the constraints\footnote{See references \cite{Hall1996,Beig2004,Kato1975} for additional details.}. 

The geometrical meaning of the covectors introduced in the previous paragraph is as follows. At a spacetime point $p\in \mathcal{M}$, the collection of all covectors satisfying condition (2) is denoted by $S_p$. This set defines a nonempty open, convex cone at $p$ and the tangent vectors $p^{a}\in T_{p}\mathcal{M}$ such that $p^{a}n_{a}>0$ for all $n_{a}\in S_p$ determine the cone of influence of the physical field, i.e., the maximal speed of propagation in any given direction. It turns out that the latter is also a nonempty open, convex cone at $p$.

%Once a particular hyperbolization is obtained, a well-known theorem guarantees that initial data given on a hypersurface $\Sigma$ with normal covector $n_{m}$ are then uniquely evolved away from the hypersurface \cite{Hall1996, Beig2004,Kato1975}.\\

In the next subsection, we start by showing that a family of symmetrizers parametrized by a vector field always exists for the system of first order PDEs given by Eqs. (\ref{EOM}). It is important to point out that our result is general in the sense that it does not depend on the specific form of the Lagrangian density. We then obtain the conditions
that a theory must satisfy for positive-definiteness (see condition (2) above) to hold. This is first obtained for a particular covector and henceforth generalized
%. The later is achieved 
by inspecting the characteristic varieties (dispersion relations), which are necessarily well-behaved for symmetric hyperbolic systems.
%From now on we restrict our analysis to a structure seen by photons given by \cite {Hat}, \cite{William}
%\begin{equation}
%X^{abcd}\equiv \frac{1}{2}g^{abcd}-4\gamma C^{abcd},
%\end{equation}
%where $g_{abcd}=g_{a[c}g_{bd]}$ and $C_{abcd}$ the Weyl tensor. 

\subsection{Symmetrizer}
In order to find a symmetrizer, it is convenient to work with \textit{projections}. In other words, we seek a multi-tensorial field $h^{A}_{\phantom a\alpha}$ such that the quantity $\delta\phi^{\alpha}(h^{A}_{\phantom a\alpha}K_{A \phantom a \beta}^{\phantom a m})\delta{\psi}^{\beta}$ is symmetric in $\delta\phi$ and $\delta{\psi}$. Making the identifications 
\begin{equation}
\delta\phi^{\alpha}\rightarrow A^{ab},\quad\quad\quad \delta{\psi}^{\alpha}\rightarrow B^{ab}, 
\end{equation}
where $A$ and $B$ are generic bivectors, we obtain from Eq. (\ref{K}), the relation
\begin{equation}\label{eqn}
K_{A \phantom a \beta}^{\phantom a m}\delta\psi^{\beta}=\left(\frac{1}{2}X_{a\phantom a\phantom a bc}^{\phantom a m} B^{bc} ,\ \star  B_{a}^{\phantom a m}\right).
\end{equation}
Since there is no known practical procedure to obtain a symmetrizer for an arbitrary system of PDE's, hyperbolizations are found, for a  sufficiently low number of dimensions, by solving explicitly the algebraic equations inherent to
the system defined by Eq. (\ref{KF}) and, in higher dimensions, by guessing. Let us show that 
the symmetrizer
is given by the projection
\begin{equation}\label{sym}
\delta\phi^{\alpha}h^{A}_{\phantom a\alpha}=\left(A^{a}_{\phantom a q}\ ,\  \frac{1}{2}\star X^{a}_{\phantom a qrs}A^{rs}\right)t^{q}
\end{equation}
where $t^{q}$ is an auxiliary vector field and $\star X^{a}_{\phantom a qrs}$ is the left Hodge dual as defined before. Indeed, multiplying (\ref{sym}) by (\ref{eqn}) one obtains
\begin{equation}
\delta\phi^{\alpha}(h^{A}_{\phantom a\alpha}K_{A \phantom a \beta}^{\phantom a m})\delta{\psi}^{\beta}=\frac{1}{2}\left(X_{a\phantom a\phantom a bc}^{\phantom a m}A^{a}_{\phantom a q}B^{bc}+        \star X^{a}_{\phantom a qrs}A^{rs}\star B_{a}^{\phantom a m}\right)t^{q}
\end{equation}
Now, defining
\begin{equation}
Y^{a}_{\phantom a q}\equiv X^{a}_{\phantom a qrs}A^{rs},\quad\quad\quad\star Y^{a}_{\phantom a q}\equiv \star X^{a}_{\phantom a qrs}A^{rs},
\end{equation}
and using the well known identity valid for anti-symmetric tensors
\begin{equation}
(\star Y^{aq})(\star B_{am})=-\frac{1}{2}(Y_{ln}B^{ln})\delta_{m}^{\phantom a q}+Y_{am}B^{aq},
\end{equation}
it follows that
\begin{equation}\label{exp}
\delta\phi^{\alpha}(h^{A}_{\phantom a\alpha}K_{A \phantom a \beta}^{\phantom a m})\delta{\psi}^{\beta} =\frac{1}{2}\left[X_{a\phantom a\phantom a bc}^{\phantom a m}(A^{a}_{\phantom a q}B^{bc}+A^{bc}B^{a}_{\phantom a q})-\frac{1}{2}(X_{abcd}B^{ab}A^{cd})\delta^{m}_{\phantom a q}\right]t^{q},
\end{equation}
which is obviously symmetric in $A$ and $B$ since $X_{abcd}=X_{cdab}$. Therefore, the symmetrizer itself is given by the simple expression
\begin{equation}\label{symm}
h^{A}_{\phantom a \alpha}=\frac{1}{2}(g^{a}_{\phantom a qrz},\star X^{a}_{\phantom a qrz})t^{q},
\end{equation}
It is remarkable that Eq. (\ref{symm}) is valid independently of the specific content of the tensor field $X_{abcd}$: in particular it does not matter whether the equations of motion contain quasi-linear terms or not.

The application of Eq. (\ref{symm}) to Eq. (\ref{K}), yields the object
\begin{equation}
\hat{K}_{\alpha\phantom a\phantom a\beta}^{\phantom a m}=\frac{1}{4}(g^{a}_{\phantom a qrz}X_{a\phantom a\phantom a bc}^{\phantom a m}+\star X^{a}_{\phantom a qrz}\varepsilon_{a\phantom a\phantom a bc}^{\phantom a m})t^{q},
\end{equation}
which, after straightforward algebraic manipulations, becomes
\begin{equation}\label{symmf}
\hat{K}_{\alpha\phantom a\phantom a\beta}^{\phantom a m}=-\frac{1}{4}\left(g_{q[a}X_{b]\phantom a \phantom a cd}^{\phantom a m}+g_{q[c}X_{d]\phantom a \phantom a ab}^{\phantom a m}+\delta^{m}_{\phantom a q}X_{abcd}\right)t^{q}.
\end{equation}
%\begin{equation}\label{symmf}
%\hat{K}_{\alpha\phantom a\phantom a\beta}^{\phantom a m}=\frac{1}{4}(t_{[a}X_{b]\phantom a\phantom a cd}^{\phantom a m}+t_{[c}X_{d]\phantom a\phantom a ab}^{\phantom a m}+t^{m}X_{abcd}),
%\end{equation}
Notice that this equation is indeed symmetric in the exchange of antisymmetric indices $ab\Leftrightarrow cd$ and that the auxiliary vector field $t^{q}$ remains (up to now) arbitrary, so we can use it at our disposal. This concludes the first step of our task.

%\begin{equation}
%M_{q}^{\phantom a m}=\left\{X^{\phantom a a}_{q},\ Y_{a}^{\phantom a m}\right\}+\frac{1}{2}X_{ab}Y^{ab}\delta_{q}^{\phantom a m},\quad\quad N_{q}^{\phantom a m}=-4\gamma\left[\left\{X_{qa},\ Y_{bc}\right\}C^{ambc} +\frac{1}{2} X_{ab}Y_{cd}C^{abcd}\delta_{q}^{\phantom a m}\right].
%\end{equation}\\
%with $\left\{X_{ab},\ Y_{cd}\right\}=X_{ab}Y_{cd}+Y_{ab}X_{cd}$, which are all symmetric quantities in X and Y. Note that the symmetrizer splits into a main term, which coincides with Maxwell's symmetrizer when $\gamma=0$ plus a term involving the conformal tensor.   

\subsection{Positive definitness}

Let us next investigate whether the above symmetrizer constitutes a true hyperbolization of the equations of motion. This will be achieved if we manage to find a covector field $n_{m}$ such that the characteristic matrix 
\begin{equation}\label{Principals}
\hat{K}_{\alpha\beta}(t,n)=-\frac{1}{4}[t_{[a}X_{b]\phantom a\phantom a cd}^{\phantom a m}+t_{[c}X_{d]\phantom a\phantom a ab}^{\phantom a m}+t^{m}X_{abcd}]n_{m}
\end{equation}
is positive definite, {\emph i.e.}
\begin{equation}
\delta\phi^{\alpha}\hat{K}_{\alpha\beta}\delta\phi^{\beta}>0,
\end{equation}
for all nonzero vectors $\delta\phi^{\alpha}\in\mathbb{R}^{6}$. A direct calculation gives the equivalent inequality
\begin{equation}\label{Equivalent}
-\left[t_{a}X_{b\phantom a\phantom a cd}^{\phantom a m}A^{ab}A^{cd}+\frac{1}{4}t^{m}X_{abcd}A^{ab}A^{cd}\right]n_{m}>0,  \end{equation}
where $A^{ab}$ is an arbitrary, but nonzero bivector. Notice that the latter is a fully covariant expression in the sense that it does not depend on any particular choice of coordinates. It will restrict, however, the choices of admissible covector and auxiliary vector fields. 

In order to proceed, we shall decompose $A^{ab}$ and $X_{abcd}$ with respect to the auxiliary vector field $t^{q}$. To do so, assuming that $t^{q}$ is timelike, future-directd and normalized, we write the bivector as
\begin{equation}
A_{ab}=\mathfrak{a}_{[a}t_{b]}+\varepsilon_{ab}^{\phantom a\phantom a cd}t_{c}\mathfrak{b}_{d},    
\end{equation}
with
\begin{equation}
\mathfrak{a}_{a}\equiv A_{ab}t^{b},\quad\quad \mathfrak{b}_{a}\equiv \star A_{ab}t^{b},\quad\quad\mathfrak{a}_{a}t^{a}=0\quad\quad\mathfrak{b}_{a}t^{a}=0.   
\end{equation}
Similarly, for any double $(2,2)$-form, we have\footnote{See for instance \cite{Senovilla1999} for similar decompositions.}\\
\begin{equation}\label{PQRS}
X_{abcd}=\left\{g_{abpq}\left(g_{cdrs}\mathbb{P}^{pr}+\varepsilon_{cdrs}\mathbb{Q}^{pr}\right)+\varepsilon_{abpq}\left(g_{cdrs}\mathbb{R}^{pr}+\varepsilon_{cdrs}\mathbb{S}^{pr}\right)\right\} t^{q}t^{s},
\end{equation}\\
with the 2-index tensors given by\\
\begin{equation}\label{2-index}
\mathbb{P}_{ab}\equiv X_{acbd}t^{c}t^{d},\quad \mathbb{Q}_{ab}\equiv-X_{acbd}\star t^{c}t^{d},\quad \mathbb{R}_{ab}\equiv -\star X_{acbd} t^{c}t^{d},\quad \mathbb{S}_{ab}\equiv\star X_{acbd}\star t^{c}t^{d},
\end{equation}\\
and orthogonal to $t^{q}$ by construction. In general, each of the latter has 9 independent components since the number of independent components of $X_{abcd}$ is 36. However, in the particular case in which $X_{abcd}=X_{cdab}$, the following simplified relations are valid:
%the unique %decomposition\\
%\begin{equation}
%X_{abcd}=\tilde{X}_{abcd}+\chi\varepsilon_{abcd}
%\end{equation}
%with $\tilde{X}_{a[bcd]}=0$ and $\chi$ an arbitrary scalar possibly depending on the electromagnetic invariants (the skewon). In this situation one has
\begin{equation}
\mathbb{P}_{ab}=\mathbb{P}_{ba},\quad\quad \mathbb{S}_{ab}=\mathbb{S}_{ba},\quad\quad\mathbb{Q}_{ab}=\mathbb{R}_{ba},
\end{equation}
so that $\mathbb{P}_{ab}$ and $\mathbb{S}_{ab}$ have 6 independent components each, while $\mathbb{Q}_{ab}$ (or, equivalently, $\mathbb{R}_{ab}$) have the remaining 9 components. Notice that, in the context of electrodynamics in material media, these tensors would be related to the permittivity, permeability and magneto-electric cross terms of the medium \cite{Balakin2017}.

Using the above decompositions in Eq. (\ref{Equivalent}), the new covariant inequality
\begin{equation}\label{newineq}
(n_{m}t^{m})[\mathbb{P}_{pr}\mathfrak{a}^{p}\mathfrak{a}^{r}-\mathbb{S}_{pr}\mathfrak{b}^{p}\mathfrak{b}^{r}]>2(n_{m}t^{n})\varepsilon^{mq}_{\phantom a\phantom a np}[\mathbb{R}_{qr}\mathfrak{a}^{p}\mathfrak{a}^{r}+\mathbb{S}_{qr}\mathfrak{a}^{p}\mathfrak{b}^{r}]
\end{equation}
follows, which is to be satisfied for any $3$-vectors $\mathfrak{a}^{p}$ and $\mathfrak{b}^{q}$, not vanishing simultaneously. Furthermore, since our considerations are essentially algebraic we may %focus our attention
restrict to an arbitrary point $p\in\mathcal{M}$. In order to complete our symmetric hyperbolization, it suffices to find a specific covector $n_{a}\in T^{*}_{p}\mathcal{M}$ satisfying inequality \eqref{newineq} since, if this is the case, there will be a unique connected, open and convex cone, $S_p$, containing the initial covector, and this cone will exhaust all possibilities. The natural choice here is the covector $n_{a}=g_{ab}t^{b}$, since the right-hand-side of Eq. (\ref{newineq}) will vanish. Now, since $n_{m}t^{m}>0$ for the latter, we obtain the equivalent inequalities
\begin{equation*}
\mathbb{P}_{pr}\mathfrak{a}^{p}\mathfrak{a}^{r}>0,\quad\quad\quad\mathbb{S}_{pr}\mathfrak{b}^{p}\mathfrak{b}^{r}<0,
\end{equation*}
for all nonzero vectors $\mathfrak{a}^{p},\mathfrak{b}^{q}$. In words, a symmetric hyperbolization is achieved for a covector $n_{a}$ if 
the 2-index tensors $\mathbb{P}_{ab}$ and $\mathbb{S}_{ab}$, obtained from $X_{abcd}$ and $\star X_{abcd}\star$ via contractions with $n^{a}$, satisfy the above inequalities.

For the sake of completness, let us a repeat the calculations in an adapted frame at $p$, such that 
\begin{equation}
g_{ab}(p)=\eta_{ab},\quad\quad\quad t^{q}=\delta^{q}_{\phantom a 0}.
\end{equation}
Using Eq. (\ref{PQRS}), it can be shown that $X_{abcd}$ has the following block matrix display:
\begin{equation}
X_{\alpha\beta}=\left( \begin{array}{cc}\mathbb{P}& \mathbb{Q}\\\\
\mathbb{Q}^{T}& \mathbb{S} \end{array} \right),
 \end{equation}
where $\mathbb{P}$, $\mathbb{Q}$ and $\mathbb{S}$ denote the covariant $3\times3$ matrices constructed with the corresponding tensors in the obvious way. In order to compute Eq. (\ref{Principals}) in matrix form it is convenient to define the auxiliary $3\times3$ matrices (%see \cite{Perlick2010} for a similar terminology
a similar notation was used in \cite{Perlick2010})
\[ \textbf{A}^{1} = \left( \begin{array}{ccc}
0 & 0 & 0 \\
0 & 0 & 1 \\
0 & -1 & 0 \end{array}  \right), \quad\quad  
\textbf{A}^{2} = \left( \begin{array}{ccc}
0 & 0 & -1 \\
0 & 0 & 0 \\
1 & 0 & 0 \end{array}  \right), \quad\quad \textbf{A}^{3} = \left( \begin{array}{ccc}
0 & 1 & 0 \\
-1 & 0 & 0 \\
0 & 0 & 0 \end{array}  \right).     \] \\
Taking into account that the transposition relation $(\textbf{A}^{k})^{T}=-\textbf{A}^{k}$ holds, a direct calculation gives:

%\begin{equation}
%X_{\alpha\beta}=\left( \begin{array}{cc}\mathbb{P}_{ij}& \mathbb{R}_{li}\varepsilon^{l}_{\phantom a jk}\\\\
%\varepsilon_{jk}^{\phantom a\phantom al}\mathbb{R}_{li}& \varepsilon_{ij}^{\phantom a\phantom a m}\varepsilon_{kl}^{\phantom a\phantom a n}\mathbb{S}_{mn} \end{array} \right),
 %\end{equation}

\begin{equation}\label{n0nk}
\hat{K}_{\alpha\beta}(n)=\left( \begin{array}{cc}n_{0}\mathbb{P}-n_{k}(\mathbb{Q}\textbf{A}^{k}-\textbf{A}^{k}\mathbb{Q}^{T})& n_{k}\textbf{A}^{k}\mathbb{S}\\\\
-n_{k}\mathbb{S}\textbf{A}^{k}& -n_{0}\mathbb{S} \end{array} \right).
 \end{equation}
 The above relation may be thought as a linear map from $T^{*}_{x}\mathcal{M}$ to
 the 21-dimensional space of symmetric $6\times 6$ matrices $\mbox{Sym}_{6}$ . The set of all symmetric positive definite matrices forms an open convex cone in $\mbox{Sym}_{6}$ with apex on the origin. It turns out that the image of the particular covector $n_{m}=\eta_{mn}t^{n}$ will lie inside this cone whenever
\begin{equation}\label{PS}
\mathbb{P}\succ 0,\quad\quad\quad\quad\mathbb{S}\prec 0.
\end{equation}
since $n_{m}=(1,0,0,0)$ in our frame. Therefore, if Eqs. (\ref{PS}) are satisfied pointwisely, symmetric hyperbolicity is guaranteed for the corresponding covector field. In other words, if the auxiliary vector field $t^{a}$ is vorticity-free, then
initial data given on a hypersurface $\Sigma$ with normal covectors $t_{a}$ are  uniquely evolved away from the hypersurface. 
As will be shown
in Sect. \ref{applic}, the requirements set by Eqs.\eqref{PS}
may yield  severe constraints on the Lagrangian density, the intensity of the curvature tensor or the electromagnetic field.

\subsection{Constraints}

We now discuss the remaining equations, which are left aside by the symmetrizer. To do so, we recall that, acording to Geroch's formalism, a constraint is a tensor $c^{An}$ such that
\begin{equation}
c^{An}K_{A\phantom a\alpha}^{\phantom a m}+c^{Am}K_{A\phantom a\alpha}^{\phantom a n}=0.    
\end{equation}
It is straightforward to show that, in our case, this tensor is given by
\begin{equation}\label{constraint}
c^{An}=(xg^{an},yg^{an})    
\end{equation}
where $x,y\in\mathbb{R}$. Indeed, multiplying Eq. (\ref{constraint}) by (\ref{K}) one obtains
\begin{equation}
c^{An}K_{A\phantom a\alpha}^{\phantom a m}=\frac{1}{2}(xX^{nm}_{\phantom a\phantom a\phantom a bc}+y\varepsilon^{nm}_{\phantom a\phantom a\phantom a bc})
\end{equation}
which is obviously antisymmetric in $n$ and $m$. Here, the emergence of two real numbers reveals that the vector space of constraints is actually 2-dimensional, as expected.

From the above calculation one concludes that the constraints are complete, in the sense that the number of constraint equations plus the number of evolution equations equals the number of initial equations. The constraints are integrable if the equation
\begin{equation}
c^{An}\nabla_{n}(K_{A\phantom a \beta}^{\phantom a m}\nabla_{m}\Phi^{\beta}+J_{A})=0,    
\end{equation}
%is satisfied not because of the original system Eq.  (\ref{KF}), but holds identically as an identity due the algebraic structure of the principal symbol. 
is identically satisfied solely due the algebraic structure of the principal symbol, independently of the equations (\ref{KF}) of the original system.
We leave for the reader to verify that this is indeed the case. This means that the original system of equations is equivalent to a symmetric-hyperbolic one with two additional integrable constraints.

\section{Characteristic cones}
\label{cones}
Suppose that we manage to find a symmetric hyperbolization with constraints, as described in the previous sections. We have seen that this choice, however, is far from unique. Indeed, by the continuity of Eq.(\ref{n0nk}), any \textit{small} deformation of $t_{a}$ will be such that condition (2) of symmetric hyperbolization
is satisfied.
%
%also yield a another positive definite  result. 
It turns out that the set of all admissible covectors $S_p$ in $T_{p}^{*}\mathcal{M}$ is determined by the unique connected, open, convex, positive cone containing the initial covector 
\cite{Beig2004}
. Its existence is related to the hyperbolicity of the characteristic polynomial, defined by
\begin{equation}\label{khat}
p(n)\equiv\mbox{det}\left( \hat{K}_{\alpha\beta}(n)\right)
%=\frac{1}{6!}[\alpha_{1}...\alpha_{6}][\beta_{1}...\beta_{6}]\hat{K}_{\alpha_{1}\beta_{1}}(n)\ ...\ \hat{K}_{\alpha_{6}\beta_{6}}(n),
\end{equation}
which is a homogeneous multivariate polynomial of degree $6$ in our case. Recall that, at a spacetime point, such a polynomial is called hyperbolic in a direction $t_{a}$ if $p(t_{a})>0$ and the univariate polynomial $p(u_{a}+\lambda t_{a})$ only has real roots for all covectors $u_{a}\neq t_{a}$. Now, the vanishing set of the characteristic polynomial will define an algebraic variety: the cone of characteristic conormals (or characteristic cone, for brevity). In general, it will consist of different codimension 1 sheets which may be nested, intersect along lines, or even coincide. Geometrically, hyperbolicity in the direction of $t_{a}$ is the requirement that every line parallel to $t_{a}$ intersects this algebraic variety at exactly $``6"$ points (counting multiplicities). Clearly, this condition severely constrains the topology of the characteristic cones, thus guaranteeing well-behaved propagation for small wavy excitations. In particular, it was shown by Garding \cite{Garding1959} that the closure of the connected component of $t_{a}$ in the set $\{n_{a}|p(n_{a})\neq 0\}$ is necessarily convex: the hyperbolicity cone of the polynomial.
That symmetric hyperbolicity implies the hyperbolicity of the characteristic polynomial is direct. To see this one simply observes that the equation 
\begin{equation}
\mbox{det}(\hat{K}_{\alpha\beta}(a+\lambda t))=0
\end{equation}
characterizes the eigenvalues of the quadratic form $\hat{K}_{\alpha\beta}(a_{m})$ relative to the metric $\hat{K}_{\alpha\beta}(t_{m})$ - and these eigenvalues have to be real (see \cite{Beig2004} for further details).

In order to compute the characteristic polynomial explicitly, we substitute Eq. (\ref{n0nk}) into Eq. (\ref{khat}) and use Schur's determinant identity to obtain the product of determinants
\begin{equation}
p(n)=-\mbox{det}(\mathbb{S})q(n),
\end{equation}
where
\begin{equation}\label{detq}
q(n)\equiv \mbox{det}\left(n_{0}^{2}\mathbb{P}-n_{0}n_{k}(\mathbb{Q}\textbf{A}^{k}-\textbf{A}^{k}\mathbb{Q}^{T})-n_{k}n_{l}\textbf{A}^{k}\mathbb{S}\textbf{A}^{l}\right).
\end{equation}
Notice that $q(t)$ is necessarily positive definite, since $\mathbb{P}>0$ and $\mathbb{S}<0$ for this particular covector. Interestingly, the determinant in Eq. (\ref{detq}) has been calculated several times in literature, see e.g. \cite{Perlick2010},\cite{Rivera2011}. In particular, it can be shown that it factorizes as
\begin{equation}
q(n)=(n_{m}t^{m})^{2}P(n),
\end{equation}
where
\begin{equation}
P(n)=\frac{1}{24}\varepsilon_{a_{1}a_{2}a_{3}a_{4}}\varepsilon_{b_{1}b_{2}b_{3}b_{4}}X^{a_{1}a_{2}b_{1}c_{1}}X^{c_{2}a_{3}b_{2}c_{3}}X^{c_{4}a_{4}b_{3}b_{4}}n_{c_{1}}n_{c_{2}}n_{c_{3}}n_{c_{4}}.
\end{equation}
In other words, the sixth order polynomial always reduces to a product of a quadratic polynomial and a quartic polynomial. Since the vanishing set of the quadratic polynomial gives a non-compact variety (plane) inconsistent with the constraint equations,
%it can be neglected. Then,
the characteristic cone is given by the
covectors $k_a$ that satisfy the
fourth-order \textit{Fresnel equation} 
\begin{equation}\label{X3}
P(k)\sim\left(\star X_{pq}^{\phantom a\phantom a ar}X^{bpcs}X^{dq}_{\phantom a\phantom a rs}\star\right)k_{a}k_{b}k_{c}k_{d}=0,
\end{equation}
The fourth rank tensor defined by the terms between parentheses is called the Kummer tensor, whereas its totally symmetrized version is usually called the Tamm-Rubilar tensor. If symmetric hyperbolicity holds, the above polynomial is necessarily hyperbolic in the direction of $t_{a}$, its vanishing set determining the causal structure of the theory up to a conformal factor. In specific situations, where $X_{abcd}$ is sufficiently simple, Eq. (\ref{X3}) will reduce to the more familiar product of quadratic polynomials
\footnote{A general \emph{Ansatz} leading to bi-metricity in nonlinear electromagnetism was presented in \cite{Visser2002}.
}
\begin{equation}\label{factorized}
P(k)\sim\left(g^{ab}_{(1)}k_{a}k_{b}\right)\left(g^{ab}_{(2)}k_{c}k_{d}\right)=0. 
\end{equation}
In these cases, symmetric hyperbolicity guarantees that the rank-2 contravariant tensors $g^{ab}_{(1)}$ and $g^{ab}_{(2)}$ are nondegenerate, necessarily of Lorentzian type and with the same signature. Furthermore, if these tensors coincide
\footnote{Covariant conditions
on the Fresnel surface for birefringence to be absent were derived in \cite{Itin2005}.
}, then 
%technical reasons constrain us to consider 
only one of them must be considered in the dispersion relation (reduced polynomial) , {\emph i.e.},
\begin{equation}
P_{red}(k)\sim g^{ab}_{(1)}k_{a}k_{b}=0.    \end{equation}
 It is important to point out that the abovementioned result stating that symmetric hyperbolicity leads to a cone strongly constraints the possible shape of the Fresnel surfaces, defined in a convenient 3-space: they must 
 be topologically equivalent to those obtained from the 4-d hyperbolicity cone. Hence, open surfaces are not allowed by symmetric-hyperbolic propagation \footnote{For examples of Fresnel surfaces in Maxwell´s theory non-minimally coupled to gravity, see \cite{Balakin2017}.
 }.  
\section{General result}
\label{result}
We have seen that symmetric hyperbolicity imposes restrictions on admissible physical theories. Importantly, they are always expressed in terms of matrix inequalities which are much easier to guarantee than to check the hyperbolicity of the corresponding characteristic polynomial. In order to obtain explicit expressions for the matrix inequalities, it is convenient to introduce the projection tensor $h_{ab}=g_{ab}-t_{a}t_{b}$, which projects arbitrary tensors onto the ``rest spaces'' orthogonal to $t^{q}$, i.e,
\begin{equation}
h_{ab}=h_{ba},\quad\quad\quad h_{ac}h^{c}_{\phantom a b}=h_{ab},\quad\quad\quad h_{ab}t^{b}=0. \end{equation}
Similarly, we decompose the electromagnetic 2-form, the Weyl tensor and the traceless part of the Ricci tensor as
\begin{equation}
F_{ab}=E_{[a}t_{b]}+\varepsilon_{abcd}t^{c}B^{d} ,  
\end{equation}
\begin{equation}\label{Weyldec}
W_{abcd}=\left\{g_{abpq}\left(g_{cdrs}\mathcal{E}^{pr}-\varepsilon_{cdrs}\mathcal{B}^{pr}\right)-\varepsilon_{abpq}\left(g_{cdrs}\mathcal{B}^{pr}+\varepsilon_{cdrs}\mathcal{E}^{pr}\right)\right\} t^{q}t^{s},
\end{equation}
\begin{equation}
S_{ab}=St_{a}t_{b}+Q_{(a}t_{b)}+N_{ab},
\end{equation}
with the following definitions
\begin{equation}\label{EMdef}
E_{a}\equiv F_{ab}t^{b},\quad\quad\quad B_{a}\equiv \star F_{ab}t^{b},
\end{equation}
\begin{equation}\label{Weyldef}
\mathcal{E}_{ab}\equiv W_{acbd}t^{c}t^{d},\quad\quad\quad \mathcal{B}_{ab}\equiv\star W_{acbd}t^{c}t^{d},    
\end{equation}
\begin{equation}\label{Riccidef}
S\equiv S_{ab}t^{a}t^{b},\quad\quad Q_{a}\equiv h_{a}^{\phantom a b}S_{bc}t^{c},\quad\quad N_{ab}\equiv h_{a}^{\phantom a c}h_{b}^{\phantom a d}S_{cd}.
\end{equation}
According to the latter, the tensor fields $\{E_{a}, B_{a},Q_{a},\mathcal{E}_{ab},\mathcal{B}_{ab},N_{ab}\}$ are automatically orthogonal to the auxiliary vector field.%From Eq. (\ref{Weyldec}) one easily sees that the dual may be obtained via the well-known transformations
%\begin{equation}
%\mathcal{E}\rightarrow\mathcal{H},\quad\quad\quad \mathcal{H}\rightarrow-\mathcal{E}.
%\end{equation}

In order to compute $\mathbb{P}_{ab}$ and $\mathbb{S}_{ab}$ from Eq. (\ref{XL}), the following relations involving the constitutive tensors are useful\\
\begin{equation}
{\phantom a}^{(1)}\chi_{abcd}t^{b}t^{d}=h_{ac},\quad\quad{\phantom a}^{(2)}\chi_{abcd}t^{b}t^{d}=0,
\end{equation}
\begin{equation}
{\phantom a}^{(3)}\chi_{abcd}t^{b}t^{d}=Rh_{ac},\quad\quad{\phantom a}^{(4)}\chi_{abcd}t^{b}t^{d}=0,
\end{equation}
\begin{equation}
{\phantom a}^{(5)}\chi_{abcd}t^{b}t^{d}=\mathcal{E}_{ac},\quad\quad{\phantom a}^{(6)}\chi_{abcd}t^{b}t^{d}=\mathcal{B}_{ac},
\end{equation}
\begin{equation}
{\phantom a}^{(7)}\chi_{abcd}t^{b}t^{d}=N_{ac}.
\end{equation}\\
Similarly, for the induction tensors, Eq. (\ref{inddef}), one obtains\\
\begin{equation}
{\phantom a}^{(1)}H_{ab}t^{b}=E_{a},\quad\quad{\phantom a}^{(2)}H_{ab}t^{b}=B_{a},
\end{equation}
\begin{equation}
{\phantom a}^{(3)}H_{ab}t^{b}=RE_{a},\quad\quad{\phantom a}^{(4)}H_{ab}t^{b}=RB_{a},
\end{equation}
\begin{equation}
{\phantom a}^{(5)}H_{ab}t^{b}=\mathcal{E}_{ab}E^{b}-\mathcal{B}_{ab}B^{b},\quad\quad{\phantom a}^{(6)}H_{ab}t^{b}=\mathcal{E}_{ab}B^{b}+\mathcal{B}_{ab}E^{b},
\end{equation}
\begin{equation}
{\phantom a}^{(7)}H_{ab}t^{b}=SE_{a}+N_{ab}E^{b}-\varepsilon_{abcd}t^{b}Q^{c}E^{d},
\end{equation}
\\

Using the above relations and the Ruse-Lanczos identities (see the Appendix A), it can be checked that the tensors $\mathbb{P}_{ab}$ and $\mathbb{S}_{ab}$ are given by:\\
\begin{eqnarray}
\label{pab}
\mathbb{P}_{ab} & = & (\mathcal{L}_1+\mathcal{L}_3R)h_{ab}
+
\mathcal{L}_{5}\mathcal{E}_{ab}+
\mathcal{L}_{6}\mathcal{B}_{ab}
+\mathcal{L}_{7}N_{ab}
\\ \nonumber
& & 
+4E_aE_b(\mathcal{L}_{11}+2R\mathcal{L}_{13})
\\ \nonumber
& &
+4 E_{(a}B_{b)} \left[
\mathcal{L}_{12}+R(\mathcal{L}_{14}+\mathcal{L}_{23})\right] 
\\ \nonumber
& & 
+4B_aB_b(\mathcal{L}_{22}+2R\mathcal{L}_{24})
\\ \nonumber
& & 
+4[\mathcal{L}_{15}
E_{(a}+\mathcal{L}_{25}B_{(a}][\mathcal{E}_{b)c}E^{c}-\mathcal{B}_{b)c}B^{c}]
\\ \nonumber
& & 
+4[\mathcal{L}_{16}
E_{(a}+\mathcal{L}_{26}B_{(a}][\mathcal{E}_{b)c}B^{c}+\mathcal{B}_{b)c}E^{c}]
\\ \nonumber
& &
+4[\mathcal{L}_{17}
E_{(a}+\mathcal{L}_{27}B_{(a}][SE_{b)}+N_{b)c}E^{c}-\varepsilon_{b)pqr}Q^{p}t^{q}E^{r}]+...
\end{eqnarray}

%\begin{eqnarray}
%\label{pab}
%\mathbb{P}_{ab} & = & (\mathcal{L}_1+\mathcal{L}_3R)h_{ab}
%+
%\mathcal{L}_{5}\mathcal{E}_{ab}+
%\mathcal{L}_{6}\mathcal{H}_{ab}
%+\mathcal{L}_{7}
%\left[g_{ab}S_{cd}t^ct^d
%-t_{(a}S^c_{\;b)}t_c-S_{ab}
%\right]
%\\ \nonumber
%& & 
%+4E_aE_b(\mathcal{L}_{11}+2R\mathcal{L}_{13})
%\\ \nonumber
%& &
%+4 E_{(a}H_{b)} \left[
%\mathcal{L}_{12}+R(\mathcal{L}_{14}+\mathcal{L}_{23})\right] 
%\\ \nonumber
%& & 
%+4H_aH_b(\mathcal{L}_{22}+2R\mathcal{L}_{24})
%\\ \nonumber
%& & 
%+2\mathcal{L}_{15}
%E_{(a}W_{b)drs}F^{rs}t^d
%\\ \nonumber
%& & 
%+2\mathcal{L}_{16}
%E_{(a}\star W_{b)drs}F^{rs}t^d
%\\ \nonumber
%& & 
%+4\mathcal{L}_{17}E_{(a}\left[
%F_{b)}^{\;n}t_dS^d_{\;n}+E^mS_{b)m}\right]
%\\ \nonumber
%& & 
%+2\mathcal{L}_{25}
%H_{(a}W_{b)drs}F^{rs}t^d
%\\ \nonumber
%& & 
%+2\mathcal{L}_{26}
%H_{(a}\star W_{b)drs}F^{rs}t^d
%\\ \nonumber
%& & 
%+4\mathcal{L}_{27}H_{(a}\left[F_{b)}^{\;e}t_dS^d_{\;e}+E^{m}S_{b)m}\right]+...
%\end{eqnarray}

\begin{eqnarray}
\label{sab}
\mathbb{S}_{ab} &  = & -(\mathcal{L}_{1}+\mathcal{L}_3
R)h_{ab}
-\mathcal{L}_5\mathcal{E}_{ab}-\mathcal{L}_6\mathcal{B}_{ab}+\mathcal{L}_7N_{ab} \\ \nonumber 
& & +4B_aB_b(\mathcal{L}_{11}+2R\mathcal{L}_{13})
\\ \nonumber
& &-4E_{(a}B_{b)}[\mathcal{L}_{12}
+R(\mathcal{L}_{14}+\mathcal{L}_{23}  )]
\\ \nonumber
& & +4E_{a}E_{b}(\mathcal{L}_{22}
+2R\mathcal{L}_{24})
\\ \nonumber
& & 
+4[\mathcal{L}_{15}
B_{(a}-\mathcal{L}_{25}E_{(a}][\mathcal{E}_{b)c}B^{c}+\mathcal{B}_{b)c}E^{c}]
\\ \nonumber
& & 
-4[\mathcal{L}_{16}
B_{(a}-\mathcal{L}_{26}E_{(a}][\mathcal{E}_{b)c}E^{c}-\mathcal{B}_{b)c}B^{c}]
\\ \nonumber
& &
-4[\mathcal{L}_{17}
B_{(a}-\mathcal{L}_{27}E_{(a}][SB_{b)}+N_{b)c}B^{c}-\varepsilon_{b)pqr}Q^{p}t^{q}B^{r}]+...
\end{eqnarray}\\
where the dots stand for possible nonlinear terms in the irreducible parts of the curvature tensor, which we shall not discuss in this work. The tensors $\mathbb{P}_{ab}$ and
$\mathbb{S}_{ab}$
given above are the most general ones when dealing with symmetric hyperbolicity in 
models of nonlinear electromagnetism coupled linearly to curvature. %However, even in these simplified scenarios,
The inequalities given in Eqs. (\ref{PS}), defined in terms of such tensors, lead to drastic restrictions on admissible theories: there must be a strong compromise between the electromagnetic quantities $\{E^{a},B^{a}\}$, the spacetime curvature expressed via $\{\mathcal{E}_{ab},\mathcal{B}_{ab},Q_{a},N_{ab}\}$ and the partial derivatives of the Lagrangian density.

%The important lesson to be learned here is that symmetric-hypebolicity will, in general, constrain the values of admissible fields for a given lagrangian density. 

%We shall analyze next several examples, 
%In what follows, our routine of analysis is the following:
%according to the following steps:
%
%\begin{itemize}
%\item {Identify the corresponding double symmetric $(2,2)$ form $X_{abcd}$ from the given Lagrangian;}
%\item{ Compute the 2-index tensors $\{\mathbb{P}_{ab},\mathbb{S}_{ab}\}$ and impose the inequalities given in Eqs. (\ref{PS});}
%\item{Calculate the dispersion relation and identify the cone of hyperbolicity.}
%\end{itemize}
%
%\subsection{General case:  nonlinear EM with linear coupling to the curvature.}
\section{Applications}
\label{applic}
Let us present next several examples that illustrate the restrictions obtained so far. For the sake of completeness, we display, in each case, the tensor $X_{abcd}$, the relevant inequalities and the corresponding characteristic polynomial. A generic feature of the latter is non-factorization: in general, the characteristic varieties are described by vanishing sets of fourth-order polynomials which do not split into products of second-order polynomials. Needless to say, this fact makes it considerably difficult to guarantee hyperbolicity straight from the characteristic polynomial. In constrast with this fact, symmetric-hyperbolicity automatically implies that the characteristic varieties are necessarily well-behaved.

\subsection{Maxwell electrodynamics $\mathcal{L}=-I^{1}$}

Our first example is Maxwell's linear theory, which is the simplest possible case in our setting. Using Eq.(\ref{XL}), it follows that
\begin{equation}
X_{abcd}=-g_{abcd}.
\end{equation}
Taking the auxiliary vector field $t^{q}$ as timelike, future-directed and normalized, Eqs.\eqref{pab} and \eqref{sab} give the 2-index tensors as
\begin{equation}
\mathbb{P}_{ab}=-h_{ab},\quad\quad\quad \mathbb{S}_{ab}=h_{ab}.
\end{equation}
In a frame such that $g_{ab}=\eta_{ab}$ and $t^{q}=\delta^{q}_{\phantom a 0}$, the corresponding $3\times 3$ matrices are
\begin{equation}
\mathbb{P}_{ij}=\delta_{ij},\quad\quad\quad \mathbb{S}_{ij}=-\delta_{ij},\quad\quad\quad i,j=1,2,3. 
\end{equation}
Since they satisfy Eqs.(\ref{PS}) identically, symmetric-hyperbolicity is guaranteed for all timelike covector fields $n_{m}=\eta_{mn}t^{n}$. A direct calculation using Eq. (\ref{X3}) leads to
\begin{eqnarray*}
P(k)&\sim&(g^{ab}k_{a}k_{b})^{2}.
\end{eqnarray*}
%Since we have repeated factors, we must ``delete" 
Only one of the quadratic polinomials is needed to obtain the usual dispersion relation for linear electromagnetic waves in vacuum:
\begin{equation}
P_{red}(k)\sim g^{ab}k_{a}k_{b}=0.    
\end{equation}
Clearly, the cone of hyperbolicity at $p$, $S_p$, is the connected component of $t_{a}$ i.e., it coincides with the set of future-directed timelike covectors, as expected.

\subsection{Minimally coupled nonlinear electrodynamics}

\subsubsection{Single invariant: $\mathcal{L}=\mathcal{L}(I^{1})$}
%Let us start with the case of a single invariant and then move on to more general cases. 
If the Lagrangian density is an arbitrary function of the invariant $I^{1}$, a direct inspection of Eq. (\ref{XL}) gives
\begin{equation}
X_{abcd}=\mathcal{L}_{1}g_{abcd}+4\mathcal{L}_{11}F_{ab}F_{cd}.    
\end{equation}
From Eqs.\eqref{pab} and \eqref{sab}, we obtain the projections
\begin{eqnarray}
\mathbb{P}_{ab}=\mathcal{L}_{1}h_{ab}+4\mathcal{L}_{11}E_{a}E_{b},\quad\quad\quad\mathbb{S}_{ab}=-\mathcal{L}_{1}h_{ab}+4\mathcal{L}_{11}B_{a}B_{b}, 
\end{eqnarray}
Simple manipulations using Eq.(\ref{PS}) then yield the matrix inequalities
\begin{equation}\label{ineq1}
\mathcal{L}_{1}\delta_{ij}-4\mathcal{L}_{11}E_{i}E_{j}\prec 0,
\end{equation}
\begin{equation}\label{ineq2}
\mathcal{L}_{1}\delta_{ij}+4\mathcal{L}_{11}B_{i}B_{j}\prec 0. \end{equation}
%In particular, taking the traces,  the scalar inequalities
%\begin{equation}
%\mathcal{L}_{1}-\frac{4}{3}\mathcal{L}_{11}E^{2}<0,\quad\quad%\quad \mathcal{L}_{1}+\frac{4}{3}\mathcal{L}_{11}B^{2}<0,    
%\end{equation}
%are obtained. 
In particular, they imply the condition $\mathcal{L}_{1}< 0$, since 
the theory should be well-behaved when either the electric or magnetic field vanish. However, notice that for field intensities violating the inequalities, the propagation of small wavy excitations about the background field will be, in general, ill-posed. 

A straightforward calculation using Eq. (\ref{X3}) gives the well-known bi-metric dispersion relation
\begin{equation}
P(k)\sim\left(g^{ab}_{(1)}k_{a}k_{b}\right)\left(g^{ab}_{(2)}k_{c}k_{d}\right),    
\end{equation}
with the \textit{effective metrics} given by
\begin{equation}
g^{ab}_{(1)}\equiv \mathcal{L}_{1}^{2}g^{ab},\quad\quad\quad g^{ab}_{(2)}\equiv -(\mathcal{L}_{1}g^{ab}+4\mathcal{L}_{11}\tau^{ab}),\quad\quad\quad \tau^{ab}\equiv F^{ac}F^{b}_{\phantom a c} ,
\end{equation}
as was shown for instance in \cite{Novello1999}.
Needless to say, if Eqs. (\ref{ineq1}) and (\ref{ineq2}) hold, the
effective metrics are well-behaved, and both possess the same Lorentzian signature. However, it should be clear from our discussion that symmetric-hyperbolicity requires more than the simple Lorentzian nature of the effective metrics. 

\subsubsection{Two invariants: $\mathcal{L}(I^{1},I^{2})$}
The case of two invariants is naturally more involved. The symmetric double $(2,2)$ form becomes\\
\begin{equation}
X_{abcd}=\mathcal{L}_{1}g_{abcd}+4\{\mathcal{L}_{11}F_{ab}F_{cd}+\mathcal{L}_{12}(F_{ab}\star F_{cd}+\star F_{ab}F_{cd})+\mathcal{L}_{22}\star F_{ab}\star F_{cd}\}.
\end{equation}\\
Apart from minor details on definitions, this tensor coincides with the so-called jump tensor obtained by Obukhov and Rubilar in \cite{Obukhov2002}. Notice also that the term involving $\mathcal{L}_{2}$ in Eq. (\ref{XL}) was discarded, since it is proportional to the Bianchi identity. We then obtain the tensors
(see Eqs.\eqref{pab} and \eqref{sab})
\begin{equation}
\mathbb{P}_{ab}=\mathcal{L}_{1}h_{ab}+4\{\mathcal{L}_{11}E_{a}E_{b}+\mathcal{L}_{12}(E_{a}B_{b}+B_{a}E_{b})+\mathcal{L}_{22}B_{a}B_{b}\},   
\end{equation}
\begin{equation}
\mathbb{S}_{ab}=-\mathcal{L}_{1}h_{ab}+4\{\mathcal{L}_{11}B_{a}B_{b}-\mathcal{L}_{12}(E_{a}B_{b}+B_{a}E_{b})+\mathcal{L}_{22}E_{a}E_{b}\},    
\end{equation}\\
which, in the frame described above, lead to the inequalities
\begin{equation}
\mathcal{L}_{1}\delta_{ij}-4\{\mathcal{L}_{11}E_{i}E_{j}+\mathcal{L}_{12}(E_{i}B_{j}+B_{i}E_{j})+\mathcal{L}_{22}B_{i}B_{j}\}\prec 0,   
\end{equation}
\begin{equation}
\mathcal{L}_{1}\delta_{ij}+4\{\mathcal{L}_{11}B_{i}B_{j}-\mathcal{L}_{12}(E_{i}B_{j}+B_{i}E_{j})+\mathcal{L}_{22}E_{i}E_{j}\}\prec 0,    
\end{equation}\\
A tedious calculation using Eq. (\ref{X3}) yields the effective metrics
\begin{eqnarray}
&& g^{ab}_{(1)}=\mathcal{X}g^{ab}+(\mathcal{Y}+\sqrt{\mathcal{Y}^{2}-\mathcal{X}\mathcal{Z}})t^{ab},\\
&& g^{ab}_{(2)}=\mathcal{X}g^{ab}+(\mathcal{Y}-\sqrt{\mathcal{Y}^{2}-\mathcal{X}\mathcal{Z}})t^{ab},
\end{eqnarray}
with 
\begin{eqnarray*}
&&\mathcal{X}=\mathcal{L}_{1}^{2}+2\mathcal{L}_{1}(G-\mathcal{L}_{22}F)+(\mathcal{L}_{12}\mathcal{L}_{12}-\mathcal{L}_{11}\mathcal{L}_{22})G^{2},\\
&&\mathcal{Y}=2\mathcal{L}_{1}(\mathcal{L}_{11}+\mathcal{L}_{22})+4(\mathcal{L}_{12}\mathcal{L}_{12}-\mathcal{L}_{11}\mathcal{L}_{22})F,\\
&&\mathcal{Z}= (\mathcal{L}_{11}\mathcal{L}_{22}-\mathcal{L}_{12}\mathcal{L}_{12}),
\end{eqnarray*}
and
%\begin{equation}
%,\quad\quad\quad 
we have used the notation 
$I^{1}\equiv F$,
%\quad\quad\quad 
$I^{2}\equiv G$,
%\end{equation}
$t^{ab}\equiv F^{ac}F^{b}_{\phantom a c}$
for conciseness. This result coincides with \cite{Obukhov2002, DeLorenci2000}, with minor modifications of notation.
%In order to compute the Fresnel equation Eq. (\ref{X3}) explicitly, we write down the relevant Hodge duals as
%\begin{equation}
%\star X_{pq}^{\phantom a\phantom a ar}=-\varepsilon_{pq}^{\phantom a\phantom a ar},\quad\quad\quad X^{dq}_{\phantom a\phantom a rs}\star=-\varepsilon^{dq}_{\phantom a\phantom a rs}.    
%\end{equation}

\subsection{Nonminimally coupled nonlinear electrodynamics.}

Since the equations of motion that follow from Eq. (\ref{action}) are very general, it is convenient to consider some particular cases. Hence, in what follows we assume that the Lagrangian density takes the form
\begin{equation}\label{simple}
\mathcal{L}(I^{1},I^{2})+\alpha I^{3}+\beta I^{5}+\gamma I^{7},    
\end{equation}
where $\mathcal{L}(I^{1},I^{2})$ is an arbitrary function of the usual electromagnetic invariants, and $\alpha, \beta$ and $\gamma$ are phenomenological parameters to be determined in principle from experiments. Such a density  is still sufficiently general to include most of the relevant important models present in literature.
%: Maxwell's theory and nonlinear electrodynamics in curved spacetime, 3-parameter non-minimal couplings with gravity, and nonlinear non-minimal extensions of electrodynamics. 
%\santiago{Add refs.}

Let us examine next some particular cases that follow from Eq.\eqref{simple}.
\\

\subsubsection*{Linear non-minimal model}
The corresponding Lagrangian density is given by
\begin{equation}
\mathcal{L}=-I^{1}+\alpha I^{3}+\beta I^{5}+\gamma I^{7},
\end{equation}
since we expect to recover Maxwell's theory in the flat spacetime regime. 
This model describes several possible non-minimal couplings of linear electrodynamics with gravity, and encompasses for instance 
the modifications of Maxwell´s theory due to one-loop vacuum polarization contributions  \cite{Drummond1979}. Clearly, the corresponding equations of motion are linear and, using Eq. (\ref{XL}), one obtains the double symmetric $(2,2)$-form as\\
\begin{equation}
\label{lform}
X_{abcd}=(\alpha R-1)g_{abcd}+\beta W_{abcd}+\gamma(g_{ac}S_{bd}-g_{ad}S_{bc}+g_{bd}S_{ac}-g_{bc}S_{ad}). \end{equation}\\
Notice that the irreducible parts of the Riemann tensor contribute in different ways to the equations of motion. Using the above prescription in Eqs.\eqref{pab} and \eqref{sab}, we obtain the following $2$-index tensors\\
\begin{equation}
\label{ps}
\mathbb{P}_{ab}=(\alpha R-1)h_{ab}+\beta \mathcal{E}_{ab}+\gamma N_{ab}    \end{equation}
\begin{equation}
\label{ss}
\mathbb{S}_{ab}=(1-\alpha R)h_{ab}-\beta\mathcal{E}_{ab}+\gamma N_{ab}    \end{equation}
Let us consider next several sub-cases:
%
%where
%\begin{equation}
%\mathcal{E}_{ab}=W_{acbd}t^{c}t^{d},\quad\quad\quad \mathcal{B}_{ab}=\star W_{acbd}t^{c}t^{d}    
%\end{equation}\\
%are the electric and magnetic parts of the Weyl tensor as measured by the timelike auxiliary vector field $t^{q}$. 
%
%$$
%W_{ab}^{  \;\;cd}=2t_{[a}t^{[c}\mathcal{E}_{b]}^{\;d]}+g_{[a}^{\;[c}\mathcal{E}_{b]}^{\;d]}+\epsilon_{abmn}t^{[c}H^{d]m}t^n+\epsilon^{cdmn}t_{[a}H_{b]m}t_n.
%$$
%
\paragraph{Scalar curvature coupling $(\alpha\neq0, \beta= 0, \gamma=0)$.}
~\\
\indent This is by far the simplest type of non-minimal coupling. Indeed, since the double symmetric $(2,2)$-form in Eq.\eqref{lform} reduces to
\begin{equation}
X_{abcd}=(\alpha R-1)g_{abcd},    
\end{equation}
the relevant $2$-index tensors read
\begin{equation}
\mathbb{P}_{ab}=(\alpha R-1)h_{ab},\quad\quad\quad \mathbb{S}_{ab}=(1-\alpha R)h_{ab}.
\end{equation}
Hence, the simple matrix inequality 
\begin{equation}
(1-\alpha R)\delta_{ij}\succ 0
\end{equation}
follows. Clearly, symmetric-hyperbolicity requires that $\alpha <1/R$, which may forbid good propagation for sufficiently high curvature, for a given $\alpha$. 
Regarding the characteristic cone, Eq. (\ref{X3}) gives
\begin{equation}
P(k)\sim(g^{ab}k_{a}k_{b})^{2},
\end{equation}
which shows that in this case the dispersion relation is governed by the background metric i.e. the causal structure is not changed by the coupling.

\paragraph{Weyl coupling $(\alpha= 0, \beta\neq 0, \gamma= 0)$.}
~\\
\indent It follows from Eqs. \eqref{ps} and \eqref{ss} that 
$$\mathbb{P}_{ab}=
-\mathbb{S}_{ab}=
-h_{ab}+
\beta \mathcal{E}_{ab}.
$$
Although the ensuing inequalities need to be examined on a case by case basis, they will lead to limits on the components of the electric part of the Weyl tensor \footnote{Conversely, for a given geometry, the inequalities may furnish $\beta$-dependent limits on the region of space-time where the propagation is symmetric-hyperbolic.}.

In order to compute the dispersion relation explicitly, we first recall that the Weyl conformal tensor has only one independent Hodge dual i.e. $\star W_{abcd}=W_{abcd}\star$. A direct calculation using Eq. (\ref{X3}) gives a quartic equation of the type
\begin{equation}
P(k)\sim G^{abcd}k_{a}k_{b}k_{c}k_{d}, 
\end{equation}
with the Kummer tensor given by\\
\begin{equation}
G^{abcd}\equiv g^{ab}g^{cd}-\frac{\beta^{2}}{3}\left(W^{apbq}W^{c\phantom a d}_{\phantom a p\phantom a q}+\frac{1}{4}g^{ab}W^{cpqr}W^{d}_{\phantom a pqr}\right)+\frac{\beta^{3}}{6}\star W_{pq}^{\phantom a\phantom a ar}W^{bpcs}W^{dq}_{\phantom a\phantom a rs}\star.
\end{equation}\\
Using the following identites due to Debever and Lanczos\\
\begin{equation}
W_{apcq}W_{b\phantom a d}^{\phantom a p\phantom a q}-\star W_{apcq}\star W_{b\phantom a d}^{\phantom a p\phantom a q}=A g_{ab}g_{cd}
\end{equation}
\begin{equation}
W_{apqr}W_{b}^{\phantom a pqr}=2Ag_{ab}    
\end{equation}\\
with $A\equiv \frac{1}{8}W_{pqrs}W^{pqrs}$, we obtain the simplified expression
\begin{equation}
G^{abcd}=\left(1-\frac{\beta^{2}}{6}A\right)g^{ab}g^{cd}-\frac{\beta^{2}}{3}W^{apbq}W^{c\phantom a d}_{\phantom a p\phantom a q}+\frac{\beta^{3}}{3}W_{pq}^{\phantom a\phantom a ar}W^{bpcs}W^{dq}_{\phantom a\phantom a rs}.    
\end{equation}

\paragraph{Traceless Ricci couplings $(\alpha= 0, \beta=0, \gamma\neq 0)$.}
~\\
%\indent In this case, we have
%\begin{equation}
%X_{abcd}=-g_{abcd}+\gamma(g_{ac}S_{bd}-g_{ad}S_{bc}+g_{bd}S_{ac}-g_{bc}S_{ad}). 
%\end{equation}

The relevant matrices take the form
\begin{equation}
\mathbb{P}_{ab}=-h_{ab}+\gamma N_{ab} ,   \end{equation}
\begin{equation}
\mathbb{S}_{ab}=h_{ab}+\gamma N_{ab}   . \end{equation}
As in the previous case, the corresponding inequalities will relate the coupling constant $\gamma$ to the curvature quantities described by $N_{ab}$.
The Kummer tensor is given by
\begin{eqnarray*}
G^{abcd}&=&g^{ab}g^{cd}-\gamma g^{ab}\   ^{(7)}\chi^{cpd}_{\phantom a\phantom a\phantom a p}+\frac{\gamma^{2}}{2}\left(^{(7)}\chi^{apb}_{\phantom a\phantom a\phantom a p}\ ^{(7)}\chi^{cqd}_{\phantom a\phantom a\phantom a q}-^{(7)}\chi^{apbq}\ ^{(7)}\chi^{c\phantom a d}_{\phantom a p \phantom a q}\right)  \\
&&\quad\quad\quad\quad\quad+\frac{\gamma^{3}}{6}\ ^{(7)}\chi^{apbq}\ ^{(7)}\chi^{crds}\ ^{(7)}\chi_{prqs}.
\end{eqnarray*}

\paragraph{Non-minimally coupled EM in a cosmological background.}
~\\
Another relevant example is that of the propagation in a cosmological background described by the flat Friedman-Lem\^aitre-Robertson-Walker metric, for which the Weyl tensor is null.  
In such a case, only the terms associated to ${\cal L}_1$, ${\cal L}_3$, and ${\cal L}_7$
survive in Eqs. \eqref{pab} and \eqref{sab}. 
Using Einstein´s equations, the traceless part of the Ricci tensor is given in terms of the matter by 
$$
S_{ab}=T_{ab}-\frac T 4 g_{ab},
$$
where $T_{ab}$ is the energy-momentum tensor and $T$, its trace. In a convenient tetrad basis, in which $T_{ab}={\rm diag}\;(\rho,p,p,p) $,
it follows that
\begin{equation}
\mathbb{P}_{ij}=
-\delta_{ij}\left[1+\beta (\rho-3p)-\frac 1 2 \gamma (\rho+p)
\right],
\end{equation}
\begin{equation}
\mathbb{S}_{ij}=
\delta_{ij}\left[1+\beta (\rho-3p)+\frac 1 2 \gamma (\rho+p)
\right],
\end{equation}
Hence, the inequalities (\ref{PS}) lead to
%\begin{eqnarray}
%1+k_3(\rho-3p)
%> \frac 1 2 k_7(\rho+p)& > & 0, \\
%1+k_3(\rho-3p)+\frac 1 2 k_7(\rho+p)& > & 0.
%\end{eqnarray}
$$
1+\beta (\rho-3p)
> \frac 1 2 |\gamma|(\rho+p),
$$
which is  trivially satisfied in the case of linear EM.
The propagation will be symmetric hyperbolic if this inequaly is satisfied at all times for which the model is valid. 

\section{Conclusions}
\label{concl}

Well-posedness is a basic requirement for any field theory, and is guaranteed by symmetric hyperbolicity. We have 
obtained a general form for the symmetrizer, given in Eq. \eqref{symm}, valid for 
a general Lagrangian theory. We have shown that symmetric hyperbolicity 
leads to a set of two inequalities for the matrices $\mathbb{P}$ and $\mathbb{S}$, whose elements are determined by a given theory. Regarding the constraints, we have verified that they are integrable.

When applied to nonlinear electromagnetism linearly coupled to curvature, the matrices $\mathbb{P}$ and $\mathbb{S}$
are expressed in terms 
the fields, the Lagrangian, and its derivatives, and also of the different quantities associated to curvature. They lead to strong constraints on the relevant quantities, which were illustrated with applications to  several particular cases. The examples show that 
while in the linear theory, no constraint arises from symmetric hyperbolic propagation, 
non-linearity leads to constraints on the field intensities, and non-minimal coupling
imposes restrictions on quantities associated to curvature. 
In the general case, symmetric hyperbolicity relates  the electromagnetic quantities $\{E^{a},B^{a}\}$, the spacetime curvature expressed via $\{\mathcal{E}_{ab},\mathcal{B}_{ab},Q_{a},N_{ab}\}$ and the partial derivatives of the Lagrangian density.

The ideas presented here can be applied in other settings, such as electromagnetism in material media. We plan to return to this  problem in a future publication.

\appendix

\section{Ruse-Lanczos identity}
Let $\chi_{abcd}$ denote an arbitrary double symmetric (2,2) form at a point $p\in\mathcal{M}$. Its double Hodge dual is defined as
\begin{equation}
\star\chi_{abcd}\star=\frac{1}{4}\varepsilon_{ab}^{\phantom a\phantom apq}\varepsilon_{cd}^{\phantom a\phantom a rs}\chi_{pqrs}.    
\end{equation}
Writing the traces of $\chi_{abcd}$ as
\begin{equation}
\chi_{ab}\equiv \chi^{c}_{\phantom a acb},\quad\quad\quad \chi\equiv \chi^{a}_{\phantom a a},
\end{equation}
we may construct the symmetric trace-free tensor 
\begin{equation}
\psi_{ab}\equiv \chi_{ab}-\frac{1}{4}g_{ab}\chi.   
\end{equation}
%Then, using 
Using elementary algebraic manipulations, 
%one shows 
it can be shown that
\begin{equation}
\star\chi_{abcd}\star+\chi_{abcd}=g_{a[c}\psi_{d]b}+g_{b[d}\psi_{c]a},    
\end{equation}
which is called the Ruse-Lanczos identity \cite{Hall2004,deFelice1990}. Aplying this identity to the %$\alpha$-th
constitutive tensors results in the following useful relations
\begin{eqnarray*}
&&\star ^{(1)}\chi^{ab}_{\phantom a\phantom a cd}\star+^{(1)}\chi^{ab}_{\phantom a\phantom a cd}=0,\\
&& \star^{(2)}\chi^{ab}_{\phantom a\phantom a cd}\star+^{(2)}\chi^{ab}_{\phantom a\phantom a cd}=0,\\
&&\star ^{(3)}\chi^{ab}_{\phantom a\phantom a cd}\star+^{(3)}\chi^{ab}_{\phantom a\phantom a cd}=0,\\
&& \star^{(4)}\chi^{ab}_{\phantom a\phantom a cd}\star+^{(4)}\chi^{ab}_{\phantom a\phantom a cd}=0,\\
&&\star ^{(5)}\chi^{ab}_{\phantom a\phantom a cd}\star+^{(5)}\chi^{ab}_{\phantom a\phantom a cd}=0,\\
&& \star^{(6)}\chi^{ab}_{\phantom a\phantom a cd}\star+^{(6)}\chi^{ab}_{\phantom a\phantom a cd}=0,\\
&&\star ^{(7)}\chi^{ab}_{\phantom a\phantom a cd}\star-^{(7)}\chi^{ab}_{\phantom a\phantom a cd}=0.
\end{eqnarray*}

%\begin{thebibliography}{50}

\bibliography{biblio} %archivo.bib

\end{document}